# Prior Effective Sample Size When Borrowing on the Treatment Effect Scale

**Hongtao Zhang\*, Keaven M Anderson, Zachary Zimmer, Gregory Golm**

BARDS, Merck & Co., Inc., North Wales, Pennsylvania, U.S.A.

*email:* squallteo@gmail.com

**and**

**Aditi Sapre**

Statistics, Organon & Co., Jersey City, New Jersey, U.S.A.

**and**

**Joseph G. Ibrahim**

Department of Biostatistics, University of North Carolina at Chapel Hill, Chapel Hill, North Carolina, U.S.A.

SUMMARY: With the robust uptick in the applications of Bayesian external data borrowing, eliciting a prior distribution with the proper amount of information becomes increasingly critical. The prior effective sample size (ESS) is an intuitive and efficient measure for this purpose. The majority of ESS definitions have been proposed in the context of borrowing control information. While many Bayesian models can be naturally extended to leveraging external information on the treatment effect scale, very little attention has been directed to computing the prior ESS in this setting. In this research, we bridge this methodological gap by extending the popular ELIR ESS definition. We lay out the general framework, and derive the prior ESS for various types of endpoints and treatment effect measures. The posterior distribution and the predictive consistency property of ESS are also examined. The methods are implemented in R programs available on GitHub: https://github.com/squallteo/TrtEffESS.

This paper has been submitted for consideration for publication in *Biometrics*

# 1. Introduction

Utilizing external control information plays an increasingly important role in the design and analysis stage of a new clinical trial. The appeals can be multi-fold. From an ethical perspective, it may reduce the number of patients exposed to the control arm that is expected to underperform the active treatment arm. For clinical trial sponsors, proper use of such information can lead to a more efficient study in terms of shorter duration and reduced cost, while achieving comparable statistical power. Leveraging external safety information might also facilitate detection of a signal for excessive toxicity early in the trial.

The Bayesian philosophy is naturally suited for this task by embedding the external control information into a prior distribution, and drawing the inference from the posterior distribution updated with the concurrent clinical trial data. In addition to the classical conjugate models such as beta-binomial and normal-normal, the two most widely used classes of parametric prior models are the meta-analytical-predictive (MAP) prior approaches (Neuenschwander et al., 2010; Schmidli et al., 2014; Zhang et al., 2023) and the power prior approaches (Ibrahim et al., 2015). While their applications have mostly been limited to early phase trials (phases 1 and 2) for some time, recently there has been a robust uptick in implementing them in the late phase setting, fueled by improved understanding of the methods and FDA's complex innovative design program (Price and Scott, 2021).

Neither borrowing too much nor too little is desirable in practice. On the one hand, having an overwhelmingly informative prior could eclipse the information in the current data that are deemed more important. On the other hand, using a non- or weakly-informative prior can reduce the appeal of Bayesian methods as the results might be largely similar to the frequentist counterparts. As such, a critical step is to elicit a prior distribution for the parameter of interest, with the proper amount of information. Summary statistics and visualizations are commonly employed to describe the prior distribution in this process.

For example, a density plot of the prior distribution can be generated to show the most likely values that the parameter may take *a priori*. It can be complemented by summary statistics such as the mean/median and 95% credible interval of the prior distribution. In addition, the effective sample size (ESS) is an important metric when eliciting a prior distribution. The ESS is a scalar that quantifies the amount of information contained in the prior distribution, making itself a more intuitive measure especially for non-statisticians. In the conjugate Bayesian models, the ESS is uniquely well-defined. For example, the $Beta(a, b)$ prior for the event rate $p$ (beta-binomial model) has an ESS of $a + b$. This means the prior adds information equivalent to $a + b$ subjects with mean $a/(a + b)$. However, defining the ESS is non-trivial without conjugacy, and there are different bases on which the ESS can be calculated: variance or precision ratio (Pennello and Thompson, 2007; Hobbs et al., 2013) and information ratio (Morita et al., 2008; Neuenschwander et al., 2020a). While these definitions might be conceptually linked, they may lead to vastly different values of ESS. Therefore, the ESS should be interpreted with caution and subject to the specific definition.

Most of the aforementioned Bayesian methods and ESS definitions have been proposed in the context of leveraging external information to augment the current control arm, as external information is more likely to be available in patients taking the placebo or standard of care rather than the experimental drug. In this setting, the corresponding ESS can be interpreted as the number of subjects borrowed from external source(s) to augment the current control arm. In other applications, the information borrowing can also be conducted on the treatment effect scale. This type of borrowing is particularly useful in pediatric drug development, where the enrollment can be prohibitively challenging and the ethical appeal is even stronger than in adult trials. Information on the treatment effect observed in the adult trial(s) is often leveraged to design and analyze the pediatric trial. For most Bayesian methods originally proposed to borrow the external control data, their ideas can be seamlessly extended to incorporating external information on the treatment effect scale, such as the power prior (Psioda and Ibrahim, 2019). There is a rich collection of statistical publications focusing on considerations and applications of pediatric extrapolation, e.g., Gamalo et al. (2022) and Travis et al. (2023).

A properly calculated and interpreted prior ESS is also paramount for prior elicitation on the treatment effect scale. However, the extension of existing prior ESS definitions in this regard is less apparent and has received little attention in the literature to the best of our knowledge. We identified and filled this methodological gap to contextualize the use of external information in studies addressing important medical needs where traditional trials are challenging. Specifically, we extend the prior expected-local-information-ratio (ELIR) ESS by (Neuenschwander et al., 2020a) to properly accommodate the treatment effect scale. Other definitions of prior ESS such as Morita et al. (2008) can be derived following the same lines of derivation. In Section 2, we first review some existing definitions of ESS that are unanimously proposed in the context of borrowing control information. In Section 3, we lay out the general framework to extend the ideas to the treatment effect scale, and provide solutions corresponding to most common types of effect measures for normal and binomial endpoints. Section 4 covers the posterior inference when new data are observed. We also examine whether a desirable property termed "predictive consistency" is preserved. Lastly, concluding remarks including generalizations to time-to-event endpoints are presented in Section 5.

## 2. A Review of Effective Sample Size Methods

### 2.1 *ESS under Conjugacy*

We start with the conjugate prior models in which the prior ESS is uniquely well-defined. Table 1 lists three widely-used conjugate prior models and their respective prior ESS.

These have been extensively employed in Bayesian borrowing contexts thanks to the ease of implementation, despite being sometimes criticized for borrowing information in a fixed manner, in the sense that the extent of borrowing is unchanged regardless whether the current and external data are congruent. Moreover, the prior ESS under conjugacy serves as an important benchmark. For any general definition of ESS without conjugacy, it is highly desirable that it yields a prior ESS that is equal or close to the counterpart when a conjugate model is used.

[Table 1 about here.]

## 2.2 *Variance/Precision Ratio ESS*

One anticipated benefit of leveraging external data is that the variability (precision) of the point estimator for the parameter of interest may be reduced (improved), with the inclusion of external data. This is generally the case when the external and current data are congruent, but not guaranteed in the presence of prior-data conflict. Therefore, an intuitive way to define ESS is to anchor it on the extent of variance reduction or precision gain. Denote by $y_0$ and $y$ the collection of external and current data, respectively. For the parameter of interest, $\theta_0$ and $\theta$ are defined likewise. The prior distribution of $\theta$ given $\theta_0$ depends on the hyperparameter(s) $\eta$, that is, $p(\theta|\theta_0, \eta)$. Pennello and Thompson (2007) defined the ESS based on the variance ratio:

$$n \times \frac{Var(\theta|y)}{Var(\theta|y, y_0, \eta)},$$

where $n$ is the sample size of the current trial. The denominator is based on the posterior distribution of $\theta$ when external data $y_0$ are utilized, while the numerator is the counterpart when external data are ignored. Using a similar idea, Hobbs et al. (2013) proposed to calculate the ESS based on the posterior precision ratio:

$$n \times \left\{ \frac{Prec(\theta|y, y_0, \eta)}{Prec(\theta|y)} - 1 \right\}.$$

One common characteristic of the two definitions is that they both rely on the posterior distribution of $\theta$, hence the observed data $y$ in the current trial. Accordingly, they are usually referred to as the posterior ESS.

Using the same principle, Neuenschwander et al. (2020a) described a less well-known variance ratio ESS that relates the variance of the prior distribution to that of a point estimator for $\theta$, denoted $\hat{\theta}_N$, from a sample of size $N$:

$$N \times \frac{E_\theta\{Var(\hat{\theta}_N|\theta)\}}{Var(\theta)}.$$

Here $N$ is different from the current trial sample size $n$. The denominator is the variance of $\theta$ based on the prior distribution. The expectation taken in the numerator can depend on $\theta$, e.g., for binomial rates. This definition of ESS does not require the current trial data. The authors modified the definition slightly so that the ESS can be expressed in terms of the Fisher information which is closely related to the variance/precision. See equation (2) of Neuenschwander et al. (2020a).

### 2.3 *Information Ratio ESS*

The first formulation of the information ratio ESS was introduced in the seminal paper by Morita, Thall and Mueller (Morita et al., 2008). The definition involves the following Fisher information terms:

- $i_F(\hat{\theta}_{IU}; \theta)$: the observed Fisher information for one information unit (IU) that estimates $\theta$.
- $i_F(\theta)$: the expected Fisher information for one IU, that is,

$$i_F(\theta) = E_{\hat{\theta}_{IU}|\theta}\{i_F(\hat{\theta}_{IU}; \theta)\}.$$

- $i(p(\theta))$: the Fisher information of the prior distribution $p(\theta)$.
- $i(p_0(\theta))$: the Fisher information of an $\epsilon$-information prior $p_0(\theta)$ with a large variance.

Morita et al. (2008) defined the prior ESS as the integer $m$ that minimizes

$$|i(p_0(\bar{\theta})) + m \cdot E\{i_F(\hat{\theta}_{IU}; \bar{\theta})\} - i(p(\bar{\theta}))|,$$

where $\bar{\theta}$ is the mean of the prior distribution $p(\theta)$ and the expectation is taken over the prior-predictive distribution under $p(\theta)$. It is worth noting that the point at which the Fisher information terms are evaluated can be arbitrary in principle.

Ignoring the negligible information $i(p_0(\bar{\theta}))$, Neuenschwander et al. (2020a) showed that the ESS can be approximated by

$$ESS_{MTM} \approx \frac{i(p(\tilde{\theta}))}{i_F(\tilde{\theta})}, \tag{1}$$

where $\tilde{\theta}$ represents the mode of the prior distribution. They further proposed the expected local-information-ratio (ELIR) ESS that removes the arbitrariness with respect to the point at which the information ratio (1) should be evaluated. The $ESS_{ELIR}$ is defined as the expected value of (1) over the prior distribution of $\theta$:

$$ESS_{ELIR} = E_\theta \left\{ \frac{i(p(\theta))}{i_F(\theta)} \right\}. \tag{2}$$

Both $ESS_{MTM}$ and $ESS_{ELIR}$ yield the standard prior ESS for most one-parameter exponential families including those in Table 1. Furthermore, $ESS_{ELIR}$ possesses another desirable property termed "predictive consistency" under certain circumstances, which will be elaborated and assessed in Section 4.

## 3. Prior ESS on the Treatment Effect Scale

### 3.1 *General Considerations*

To this point, all ESS definitions reviewed were proposed in the context of borrowing external control data. Leveraging external information regarding the treatment effect, on the other hand, adds extra complexity and challenges in calculating the ESS. For example, one apparent challenge is to account for the randomization ratio in the current trial. On another note, the choice of treatment effect measure, e.g. rate difference versus odds ratio for binomial endpoints, could require different approaches to compute the ESS.

There are two critical aspects that enable our derivations. The first one is a properly identified information unit for which the denominator quantity $i_F(\theta)$ in (2) can be evaluated. This matter is trivial when leveraging external control data, as the IU is just one subject in the current control arm. Therefore, $i_F(\theta)$ is the inverse of the variance corresponding to the data from one control subject, as determined by the distribution of the endpoint (Table 1 in Neuenschwander et al. (2020a)). On the other hand, the IU for the treatment effect contains subjects assigned to both treatment and control arms. Data from the information unit are used to estimate the treatment effect $\theta$. The choice of the treatment effect measure and the randomization ratio in the IU are both expected to determine the Fisher information $i_F(\theta)$.

The other aspect is a correct interpretation of information ratio-based ESS definitions including the ELIR ESS. The ESS defined as such is essentially the multiple of information contained in the prior distribution, in terms of the information contained in the IU. We take a step back and use the example of borrowing external control data to demonstrate this point. In this context, the smallest IU contains one subject in the current control arm. If the calculated ELIR ESS is 10, the number of subjects borrowed is then 10 (10×1). Alternatively, we can elect to use an IU that consists of two control subjects. Leaving all other assumptions unchanged, this should yield an ELIR ESS of 5, suggesting the prior contains 5 IU-worth of information. The number of subjects borrowed, however, remains unchanged at 10 (5 × 2). The minimal IU of size 1 is frequently assumed implicitly in real applications. While it is unnecessary to differentiate the ESS in terms of IU and total number of subjects when only borrowing control information, acknowledging their difference becomes critical in calculation on the treatment effect scale.

The general procedures to calculate the ELIR ESS on the treatment effect scale can be summarized as follows:

(1) Identify the Fisher information of the prior distribution, $i(p(\theta))$.

(2) Identify the IU for the the treatment effect estimate, and compute its expected Fisher information $i_F(\theta)$. The IU should maintain the randomization ratio in the current trial.

(3) Compute the expected information ratio (2) as the ELIR ESS in terms of IU.

(4) Scale to obtain the ELIR ESS in terms of total subjects based on the IU size.

(5) If needed, proportion the ESS to obtain the group-wise ESS in each arm.

Assuming the current trial has two arms (treatment and control), in the remainder of this section, we apply this framework and derive the prior ELIR ESS corresponding to different types of endpoints and treatment effect measures.

### 3.2 *Normal Endpoints with Known Variance*

Suppose that the endpoint is normally distributed with known variance $s_1^2$ and $s_0^2$ within treatment and control arms, respectively. The most common treatment effect measure $\theta = \mu_1 - \mu_0$ is the difference between the means of the treatment and control arms, which is also normally distributed. A normal prior $N(\theta_0, s^2)$ for $\theta$ is usually posed in practice. With an $a$-to-$b$ (treatment to control) randomization ratio in the current trial, the smallest IU contains $a$ treatment subjects and $b$ control subjects. The treatment effect $\theta$ can be estimated by difference in the sample means

$$\frac{y_1 + \cdots + y_a}{a} - \frac{x_1 + \cdots + x_b}{b}, \tag{3}$$

where $y$ and $x$ respectively denote the observations from the treatment and control arms. The variance of the IU (3) is then

$$\sigma_{IU}^2 = \frac{s_1^2}{a} + \frac{s_0^2}{b}. \tag{4}$$

The variance (4) does not depend on $\theta$, suggesting that the expected Fisher information $i_F(\theta)$ equals to the inverse of $\sigma_{IU}^2$. On the other hand, the Fisher information of the normal prior $i(p(\theta))$ is $s^{-2}$ which does not depend on $\theta$ either. Therefore, no expectation needs to be taken in (2), and the ELIR ESS with respect to the IU (3) has a simple analytical form

$$ESS_{ELIR} = \sigma_{IU}^2/s^2. \tag{5}$$

**Example 1:** Suppose that we have a hypothetical current trial with a normal endpoint and a 2-to-1 randomization ratio. The common variance for the normal endpoints in the treatment $(s_1^2)$ and control $(s_0^2)$ arms equals 1. For the treatment effect $\theta$, we assign the normal prior $\theta \sim N(\theta_0 = 0, s^2 = 0.5^2)$. Table 2 shows the ELIR ESS in terms of IU or total number of subjects corresponding to several different sizes of IU. It can be observed that the ESS in terms of IU changes in a linear pattern as the IU is constructed with different numbers of subjects, while still maintaining the 2-to-1 randomization ratio. On the other hand, the ELIR ESS in terms of total subjects is invariant to the choice of IU. The total ESS can be further proportioned according to the randomization ratio. In this example, this prior is borrowing respectively 12 and 6 patients in the treatment and control arms *a priori*.

[Table 2 about here.]

The ELIR ESS is also invariant to the prior mean $\theta_0$ in this setting. For example, the total ELIR ESS remains the same at 18 when the prior $N(\theta_0 = 1, s^2 = 0.5^2)$ is assigned for $\theta$.

We conclude this subsection by pointing out that a normal endpoint with known variance is the easiest scenario to deal with for two reasons. First, the IU (4) to estimate $\theta$ is linear. Second, neither $i_F(\theta)$ nor $i(p(\theta))$ depends on $\theta$. The calculation will become more complex for binomial endpoints when one or both conditions are violated.

### 3.3 *Binomial Endpoints*

Denote the event rates in the treatment and control arms by $p_1$ and $p_0$, respectively. Regardless of the treatment effect measure, the information unit for estimating $\theta$ is calculated from the data of $a$ treatment and $b$ control subjects in a current trial with an $a$-to-$b$ randomization ratio. We derive the ELIR ESS for three treatment effect measures for binomial endpoints: risk difference, odds ratio and rate ratio (Miettinen and Nurminen, 1985).

3.3.1 *Risk Difference.* We start with the risk difference (RD), $\theta = p_1 - p_0$. The ELIR ESS for RD is the easiest of the three because the information unit can be constructed in a linear manner that is largely reminiscent of (3):

$$\frac{y_1 + \cdots + y_a}{a} - \frac{x_1 + \cdots + x_b}{b},$$

where $y$ and $x$ are random Bernoulli observations from the treatment and control arms. The IU is a consistent estimator of $\theta$ and asymptotically normal. Unlike the case of normal endpoints, its variance now depends on $p_1$ and $p_0$, and thus $\theta$. The notation is updated to properly reflect this characteristic

$$\sigma_{IU}^2(\theta(p_0, p_1)) = \frac{p_1(1 - p_1)}{a} + \frac{p_0(1 - p_0)}{b}. \tag{6}$$

This dependence on $\theta$ introduces extra complexity in the computation as the expectation of $i_F(\theta(p_0, p_1))$ needs to be taken with respect to the joint distribution of $(p_0, p_1)$, which suggests a univariate prior on $\theta$ would be insufficient. Therefore, we propose to assign a joint bivariate normal prior for the transformed parameters $l_0 = \log(p_0/(1 - p_0))$ and $\theta$:

$$\begin{pmatrix} l_0 \\ \theta \end{pmatrix} \sim N_2 \left( \begin{pmatrix} \mu_0 \\ \theta_0 \end{pmatrix}, \begin{bmatrix} m_0^2 & \rho m_0 s \\ \rho m_0 s & s^2 \end{bmatrix} \right). \tag{7}$$

There are several advantages to parameterize the prior distribution as such. First, it is more natural to assume normality on the logit scale of $p_0$. Second, marginally, we still have a univariate normal prior for $\theta \sim N(\theta_0, s^2)$ which is easy to work with. Finally, the joint distribution of $(p_0, p_1)$ can be conveniently deduced using the change-of-variable technique.

The ELIR ESS no longer has the analytical form (5), and the apparent challenge is to compute the expectation of the numerator

$$ESS_{ELIR} = \frac{E_{\theta(p_0, p_1)} \{ \sigma_{IU}^2(\theta(p_0, p_1)) \}}{s^2}. \tag{8}$$

This expectation in the numerator can be expressed in the form of a double integral

$$\int_0^1 \int_0^1 \left[ \frac{p_1(1-p_1)}{a} + \frac{p_0(1-p_0)}{b} \right] f(p_0, p_1) dp_0 dp_1,$$

where $f(p_0, p_1)$ is the joint density of $(p_0, p_1)$ whose closed form is provided in the Appendices. We evaluate this double integral using Gaussian quadrature over a fine grid in the $[0, 1] \times [0, 1]$ plane.

One practical challenge is to assign values to the hyperparameters in the joint bivariate normal prior (7). In particular, the choice of $\rho$ can be difficult as it does not have an intuitive interpretation due to the parameterization. One possible solution is to use external data to guide the prior elicitation. Based on the re-parameterization $(l_0, \theta)$ of the binomial likelihood, we derived the gradient and Hessian matrix and implemented the Newton-Raphson algorithm in our companion R programs/package to guide the prior elicitation of $\rho$. Suppose that in the external data, we observe 20 events out of 100 subjects in the control group, and 70 out of 200 in the treatment group. The point estimates for $(l_0, \theta)$ are $(-1.386, 0.15)$. Its covariance matrix is

$$\begin{bmatrix} 0.0625 & -0.01 \\ -0.01 & 0.00274 \end{bmatrix},$$

which implies $\rho$ is estimated to be -0.765.

***Example 2:*** Consider a current trial with a 2-to-1 randomization ratio, in which the treatment effect is measured by the risk difference. For the bivariate normal prior, we set $\rho = -0.8$. Marginally, the logit of $p_0$ has a weakly informative prior $N(\mu_0 = -1, m_0^2 = 1^2)$. We compute the ELIR ESS corresponding two marginal priors of $\theta$ with identical variance but different means: $N(\theta_0 = 0.3, s^2 = 0.1^2)$ and $N(\theta_0 = 0.4, s^2 = 0.1^2)$. The induced joint distributions of $(p_0, p_1)$ are illustrated in Figure 1. The two priors are moderately informative. Information units of various sizes are also examined.

[Figure 1 about here.]

[Table 3 about here.]

The results are presented in Table 3. Given $\theta_0$, the ELIR ESS behaves largely similar to that of the normal endpoints: the ESS in terms IU changes linearly as the IU scales, but the ESS in total number of subjects remains the same. The difference is that the ELIR ESS is no longer invariant to $\theta_0$ as is the case for normal endpoints. This is because the expectation term $E_\theta \{\sigma^2_{IU}(\theta(p_1, p_0))\}$ depends on $\theta_0$.

3.3.2 *Odds Ratio.*    Another extensively used treatment effect measure is the odds ratio (OR). The conventional practice is to work on the log-OR scale, where

$$\theta = \log \left\{ \frac{p_1/(1 - p_1)}{p_0/(1 - p_0)} \right\}.$$

In this case, it is no longer possible to construct a linear IU like (3) that estimates $\theta$. However, we can still apply the general procedures in Section 3.1 to derive the ELIR ESS for log-OR.

The key step is to ascertain the Fisher information $i_F(\theta)$ for an IU, which takes the form of the sample log-OR in this context: $\widehat{\theta}_{IU} = \log \left\{ \frac{\hat{p}_1(1 - \hat{p}_0)}{\hat{p}_0(1 - \hat{p}_1)} \right\}$. It consistently estimates $\theta$ and is asymptotically normal with variance

$$\sigma^2_{IU}(\hat{p}_0, \hat{p}_1) = \frac{1}{a \cdot \hat{p}_1(1 - \hat{p}_1)} + \frac{1}{b \cdot \hat{p}_0(1 - \hat{p}_0)}. \tag{9}$$

It is immediately recognized that (9) is a random variable as it is a function of $(\hat{p}_0, \hat{p}_1)$ rather than $(p_0, p_1)$. The non-linear nature of IU is attributable to this complication. Deriving the distribution of $1/\sigma^2_{IU}(\hat{p}_0, \hat{p}_1)$ and then computing the expectation as $i_F(\theta)$ can be extremely challenging. Alternatively, one may adopt a simulation-based approach that generates random binomial random variables given probabilities $p_0$ and $p_1$. The expectation of $1/\sigma^2_{IU}(\hat{p}_0, \hat{p}_1)$ can then be approximated by the empirical mean. Although plausible, this inevitably makes the computation burdensome.

We propose to replace the sample proportions with the true proportions in (9) such that

$$\sigma^2_{IU}(\theta(p_0, p_1)) = \frac{1}{a \cdot p_1(1 - p_1)} + \frac{1}{b \cdot p_0(1 - p_0)}. \tag{10}$$

Similar to the case of risk difference, we assign a joint bivariate normal prior in the same form of (7), for $l_0 = log(\frac{p_0}{1-p_0})$ and $\theta = log(\frac{p_1}{1-p_1}) - log(\frac{p_0}{1-p_0})$. The ELIR ESS in terms of number of IUs can be computed using the same equation (8). With log-OR, the expectation term of the numerator takes the form

$$\int_0^1 \int_0^1 \left[ \frac{1}{a \cdot p_1(1-p_1)} + \frac{1}{b \cdot p_0(1-p_0)} \right] f(p_0, p_1) dp_0 dp_1,$$

where $f(p_0, p_1)$ is the induced joint density of $(p_0, p_1)$. The double integral again can be efficiently evaluated by Gaussian quadrature. We also developed the Newton-Raphson algorithm to guide the prior elicitation.

It is apparent that working with (10) rather than (9) is computationally more efficient as it eliminates the need to approximate $E\{1/\sigma_{IU}^2(\hat{p}_0, \hat{p}_1)\}$ via simulations in the Gaussian quadrature. Moreover, it has an appeal from an asymptotic perspective. Recall that we can choose to use IUs of different sizes in computing the ELIR ESS, provided that the randomization ratio is maintained. As the IU gets sufficiently large, the sample proportions should converge to the true probabilities. In this sense, the asymptotic ELIR ESS is computed by using (10).

***Example 3:*** Assume the current trial has a 2-to-1 randomization ratio. We compute the ELIR ESS corresponding to two different marginal priors for log-OR $\theta$, namely, $N(\theta_0 = 0.5, s^2 = 0.5^2)$ and $N(\theta_0 = 0, s^2 = 1^2)$. The induced joint distributions of $(p_0, p_1)$ are illustrated in Figure 2. The other parameters are fixed: $l_0$ marginally has a weakly informative normal prior $N(\mu_0 = -1, m_0^2 = 0.5^2)$, and the correlation coefficient $\rho$ is set to $-0.8$. The second prior should be less informative, as it has a larger $s^2$. It is also favoring the no treatment effect scenario (log-OR= 0), as it is centered near the diagonal line of $p_0 = p_1$. Three sizes of IU are examined. The prior ELIR ESS is presented in Table 4. Similar to the desirable trend observed for the risk difference, the ESS in number of IUs changes linearly with the IU size, while the ESS in total subjects is invariant. Further results also affirm that the second prior is less informative with a smaller ELIR ESS (25.29 versus 92.92). In results not presented here, we explore the ELIR ESS computed by a simulation-based approach. With a sufficiently large IU, e.g., 10000 treatment and 5000 control subjects, the ELIR ESS approaches the corresponding values reported in Table 4.

[Figure 2 about here.]

[Table 4 about here.]

3.3.3 *Risk Ratio.* The third treatment effect measure for binomial endpoints is the risk ratio (RR) $\theta = p_1/p_0$. Without providing details, we point out that its ELIR ESS can be derived in a similar manner with log-OR. To construct an IU, according to Katz et al. (1978), the point estimator for $\theta$ is given by $\hat{\theta} = \hat{p}_1/\hat{p}_0$. Its log-transformation is consistent and asymptotically normal with variance

$$\sigma^2_{IU} = \frac{1 - \hat{p}_1}{a \cdot \hat{p}_1} + \frac{1 - \hat{p}_0}{b \cdot \hat{p}_0}.$$

By posing a joint bivariate normal prior (7) for $l_0 = log(\frac{p_0}{1-p_0})$ and $\theta = \log(p_1) - \log(p_0)$, the same procedure used for log-OR can be followed to compute the ELIR ESS.

## 4. Posterior ESS and Predictive Consistency Check

The methods articulated Section 3 concern the ESS for a prior distribution of the parameters. After the data in the new trial is collected, it is used to update the prior to obtain a posterior distribution for which a posterior ELIR ESS can in turn be computed.

We first derive a closed form of the posterior distributions when a bivariate normal prior is used for $\boldsymbol{\nu} = (l_0, \theta)^T$ concerning treatment effect measures for binomial endpoints. Specifically, denote the bivariate prior

$$\boldsymbol{\nu} \sim N_2\left(\boldsymbol{\omega_0} = \begin{pmatrix} \mu_0 \\ \theta_0 \end{pmatrix}, \Sigma_0 = \begin{bmatrix} m_0^2 & \rho m_0 s \\ \rho m_0 s & s^2 \end{bmatrix}\right).$$

With the observed new data, the sampling distribution of the parameter estimates also follows a bivariate normal distribution asymptotically, that is,

$$\widehat{\boldsymbol{\nu}}|\boldsymbol{\nu} \sim N_2(\boldsymbol{\nu}, \widehat{\Sigma}),$$

where $\widehat{\boldsymbol{\nu}}$ and $\widehat{\Sigma}$ can be obtained using the Newton-Raphson algorithm. The posterior density of $\boldsymbol{\nu}$ is then

$$p(\boldsymbol{\nu}|\widehat{\boldsymbol{\nu}}) \propto p(\widehat{\boldsymbol{\nu}}|\boldsymbol{\nu}) \cdot p(\boldsymbol{\nu})$$
$$\propto \exp\left\{-\frac{1}{2}\left[\boldsymbol{\nu}^t(\widehat{\Sigma}^{-1} + \Sigma_0^{-1})\boldsymbol{\nu} - 2\boldsymbol{\nu}^t(\widehat{\Sigma}^{-1}\widehat{\boldsymbol{\nu}} + \Sigma_0^{-1}\boldsymbol{\omega_0})\right]\right\}.$$

After completing the square in the quadratic forms in the above expression, we have

$$p(\boldsymbol{\nu}|\widehat{\boldsymbol{\nu}}) \propto \exp\left\{-\frac{1}{2}\left[(\boldsymbol{\nu} - \boldsymbol{M}^{-1}\boldsymbol{b})^t\boldsymbol{M}(\boldsymbol{\nu} - \boldsymbol{M}^{-1}\boldsymbol{b}) - \boldsymbol{b}^t\boldsymbol{M}^{-1}\boldsymbol{b}\right]\right\},$$

where

$$\boldsymbol{M} = \widehat{\Sigma}^{-1} + \Sigma_0^{-1} \text{ and } \boldsymbol{b} = \widehat{\Sigma}^{-1}\widehat{\boldsymbol{\nu}} + \Sigma_0^{-1}\boldsymbol{\omega_0}.$$

It can thus be recognized that the posterior distribution of $\boldsymbol{\nu}$ is still bivariate normal,

$$\boldsymbol{\nu}|\widehat{\boldsymbol{\nu}} \sim N_2(\boldsymbol{M}^{-1}\boldsymbol{b}, \boldsymbol{M}^{-1}). \tag{11}$$

For the univariate case of Section 3.2, the posterior distribution reduces to

$$\theta|\widehat{\theta} \sim N\left(\frac{\theta_0/s^2 + \widehat{\theta}/\sigma^2}{1/s^2 + 1/\sigma^2}, \left(\frac{1}{\sigma^2} + \frac{1}{s^2}\right)^{-1}\right), \tag{12}$$

where $\widehat{\theta}|\theta \sim N(\theta, \sigma^2 = s_1^2/n_1 + s_0^2/n_0)$ is obtained using new data from $n_0$ $(n_1)$ subjects in control (treatment) arm of the current trial.

Since the posterior distribution is also (bivariate) normal, the posterior ELIR ESS can be computed using the same methods in Section 3. More importantly, we are in a position to examine whether the desirable "predictive consistency" property is preserved when the borrowing is conducted on the treatment effect. Per the definition in Neuenschwander et al. (2020a), predictive consistency requires that for a sample of size $N$, the expected posterior ESS must be the sum of the prior ESS and $N$. It was shown in that paper that the ELIR ESS possesses this property for one-parameter exponential families, in the context of borrowing external control data. Here, we additionally explore this property when borrowing is performed on the treatment effect scale.

***Example 1 (continued):*** We first examine the case when the endpoints are normally distributed with known variance. Suppose that the new trial enrolled 200 and 100 subjects, respectively, in the treatment and control arms. If the IU contains two treatment and one control subjects, the 300 subjects in the new trial is equivalent to 100 IUs.

Regardless of the number of events observed in each arm, the sample mean difference $\widehat{\theta}$ has a variance of $\sigma^2 = 1/200 + 1/100 = 0.015$. The variance of the posterior normal distribution can be computed according to (12) by plugging in $\sigma^2 = 0.015$ and $s = 0.5$. The variance of IU, $\sigma_{IU}^2$, is unchanged at 1.5. The resulting posterior ELIR ESS is then 106 IUs, or 318 subjects. Recall that their counterparts (row 1 in Table 2) for the prior distribution are 6 IUs, or 18 subjects. It is clear that the predictive consistency property is preserved on both the IU and subject level.

This should be unsurprising as the formulation in Section 3.2 can be viewed as a special case of a normal endpoint in Neuenschwander et al. (2020a) and the posterior variance $\sigma^2$ does not depend on the sample means.

***Examples 2 and 3 (continued):*** Checking the predictive consistency in the case of binomial endpoints is more involved and requires simulations, since $\sigma_{IU}^2$ depends on $p_0$ and $p_1$, and thus $\theta$. We assume that the true event rates are 0.65 and 0.4, respectively, in the current treatment and control arms. The following steps are carried out:

(1) Simulate the random binomial variable in each arm.

(2) Obtain the point estimate and asymptotic covariance matrix for the parameters $(l_0, \theta)^T$ using the Newton-Raphson algorithm.

(3) Update and obtain the posterior distribution according to (11).

(4) Calculate the posterior ELIR ESS.

For a given sample size in the new trial, these steps are repeated 100 times. We report the prior ESS, the average posterior ESS and the quantity "average posterior ESS - prior ESS - current trial sample size". If predictive consistency is achieved, the last quantity should be equal or close to 0.

Predictive consistency is checked for the prior with $\theta_0 = 0.4$ in Example 2 (risk difference). Four different current trial sample sizes are considered. The results are reported in the upper half of Table 5. The average posterior ESS is well-behaved in the sense that it increases in a similar magnitude with the current trial sample size. However, it appears that the predictive consistency property is not achieved, as the last column is not near 0. As a positive note, this value indeed tends to stabilize and becomes increasingly negligible, as the current trial sample size gets large. We also performed the check for the prior with $\theta_0 = 0.4, s = 0.5$ in Example 3 (odds ratio). The same observations are made in the lower half of Table 5.

[Table 5 about here.]

The lack of predictive consistency here is again not unexpected. Neuenschwander et al. (2020a) pointed out this property does not hold universally outside one-parameter exponential families. In particular, some comments (Park and Lin, 2020) and the rejoinder (Neuenschwander et al., 2020b) to the paper concluded that achieving predictive consistency based on the ELIR ESS definition is extremely difficult in multivariate contexts like the ones we considered here. Our observations echo those statements.

## 5. Concluding Remarks

Bayesian methods are becoming increasingly popular for external data borrowing and evidence extrapolation in situations where standalone randomized trials are challenging. The ESS is a useful tool for prior elicitation which is arguably the most important step in implementing a Bayesian model. In a systematic and statistically rigorous manner, we bridged a critical methodological gap by extending the idea of ELIR ESS to the treatment effect setting. We started from the simple case of normal endpoints with known variance, and proposed solutions in response to the additional complexity associated with binomial endpoints. In particular, defining the prior ELIR ESS for the latter requires a joint bivariate prior for both the effect measure $\theta$ and the log-odds of the control rate $l_0$. This could be practice-changing as the convention has been to assign a univariate for $\theta$ alone. A careful examination of the bivariate prior is merited. Contour plots of the joint prior density like Figures 1 and 2 can be helpful for this purpose.

Caution is advised when calculating the prior ESS for the log odds-ratio of binomial rates. According to (10), the variance may become large when either $p_0$ or $p_1$ is close to 0 or 1. Therefore, if the joint prior $f(p_0, p_1)$ places a fairly high density in the vicinity of the boundary in the $[0, 1] \times [0, 1]$ plane, the numerator in (8) will be large which in turn could over-estimate the prior ESS. It also undermines the performance of some adaptive numerical integral routines, causing issues like non-convergence and extreme values. In our calculation, we used the more robust Gaussian quadrature to mitigate this problem, albeit it may not be the most efficient option. From the practical perspective, however, the aforementioned joint prior is seldom of true interest, and we again emphasize the importance of carefully examining the joint prior of $(p_0, p_1)$. The ESS for RR has a similar problem when $p_0$ or $p_1$ is close to 0. In contrast, the RD does not experience this issue. In fact, Psioda et al. (2022) advocated the use of RD over other treatment effect measures of binomial endpoints.

Further research will be to derive the prior ESS for time-to-event endpoints with right censoring. While the same general procedures in Section 3.1 may be applied, the development and computations will become much more complex. For example, when the borrowing is carried out on the log-hazard ratio scale, the Fisher information of IU $i_F(\theta)$ is determined by the number of events in each arm (Schoenfeld, 1981). Even under a simple exponential failure time model, the censoring distribution will add extra dimensions in the integral for the expectation terms. Simulation-based approaches may turn out to be the only viable solution. The authors are addressing the time-to-event endpoint in a separate ongoing project, which also covers the case of count response data.

Drafted April 19, 2024

APPENDIX CHANGE-OF-VARIABLE DERIVATIONS FOR BINOMIAL ENDPOINTS

The joint distribution of $(l_0, \theta)$ is bivariate normal

$$\begin{pmatrix} l_0 \\ \theta \end{pmatrix} \sim N_2\left( \begin{pmatrix} \mu_0 \\ \theta_0 \end{pmatrix}, \begin{bmatrix} m_0^2 & \rho m_0 s \\ \rho m_0 s & s^2 \end{bmatrix} \right).$$

The joint density is

$$f(l_0, \theta) = \frac{1}{2\pi m_0 s \sqrt{1-\rho^2}} \cdot$$
$$\exp\left\{ -\frac{1}{2(1-\rho^2)} \left[ \left( \frac{l_0 - \mu_0}{m_0} \right)^2 + \left( \frac{\theta - \theta_0}{s} \right)^2 - 2\rho \frac{(l_0 - \mu_0)(\theta - \theta_0)}{m_0 s} \right] \right\}$$

Through a change-of-variables, the induced joint density for $(p_0, p_1)$ has the general form,

$$f(p_0, p_1) = \frac{|J|}{2\pi m_0 s \sqrt{1-\rho^2}} \cdot$$
$$\exp\left\{ -\frac{1}{2(1-\rho^2)} \left[ \left( \frac{h_1(p_0, p_1) - \mu_0}{m_0} \right)^2 + \left( \frac{h_2(p_0, p_1) - \theta_0}{s} \right)^2 - \right.\right.$$
$$\left.\left. 2\rho \frac{(h_1(p_0, p_1) - \mu_0)(h_2(p_0, p_1) - \theta_0)}{m_0 s} \right] \right\},$$

where $h_1(p_0, p_1)$, $h_2(p_0, p_1)$, and the Jacobian determinant $|J|$ correspdoning to each treatment effect measure can be found in Table 6.

[Table 6 about here.]

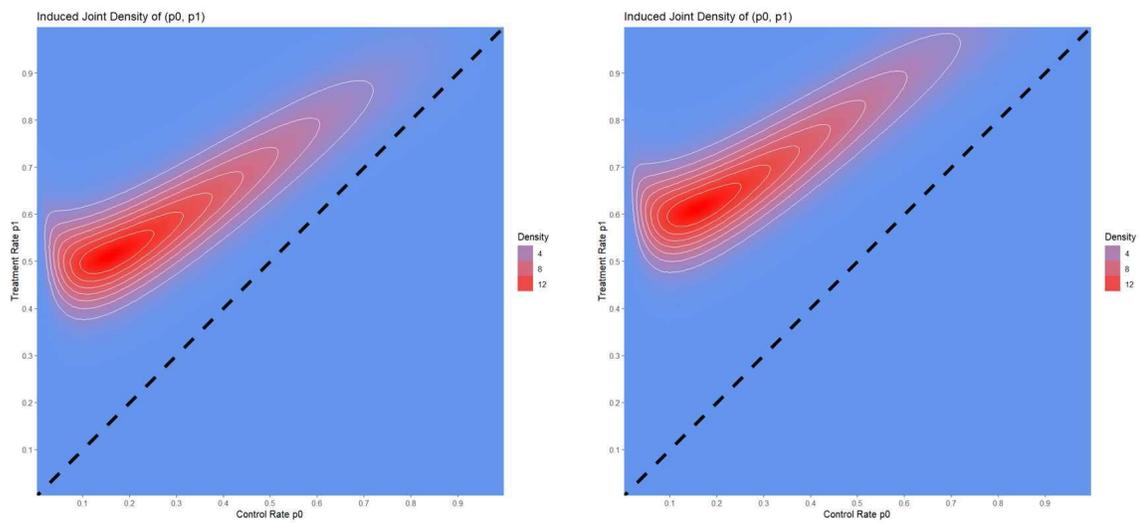

**Figure 1.**  Priors in Example 2: $\theta_0 = 0.3$ (Left); $\theta_0 = 0.4$ (Right)

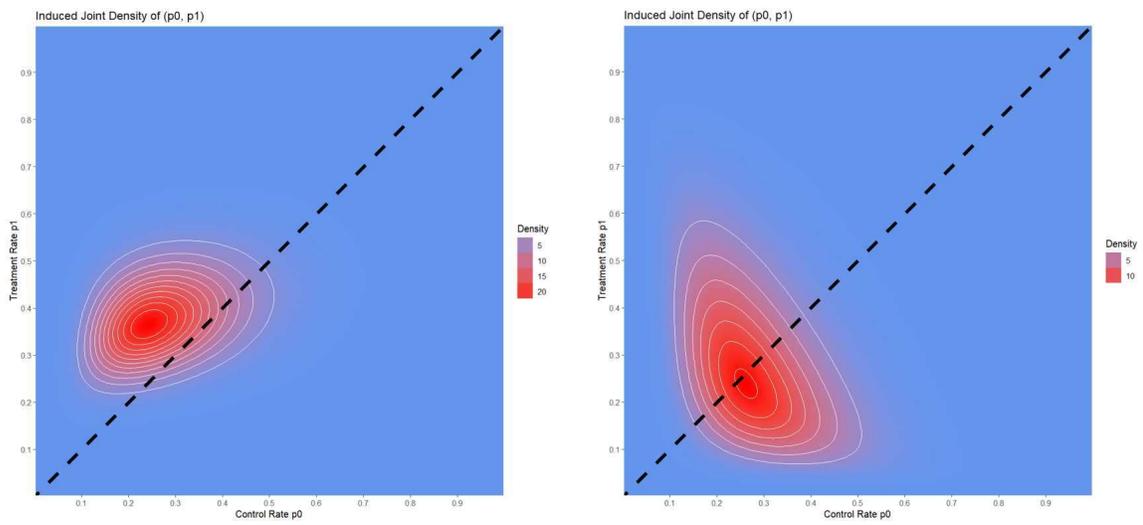

**Figure 2.** Priors in Example 2: $\theta_0 = 0.5, s = 0.5$ (Left); $\theta_0 = 0, s = 1$ (Right)

**Table 1**
*Common Conjugate Priors and Corresponding ESS*

| Endpoint Type | Parameter | Conjugate Prior | Prior ESS |
|---|---|---|---|
| Binomial | Event rate | $Beta(a, b)$ | $a + b$ |
| Normal (known variance $s^2$) | Mean | $N(\theta_0, s^2/n_0)$ | $n_0$ |
| Poisson | Incidence rate | $Gamma(a, b)^1$ | $b$ |

[1] Natural parameterization: $a$ =shape, $b$ =rate

**Table 2**
*Prior ELIR ESS for the Difference Between Two Normal Means*

| IU (Trt:Ctrl†) | $\sigma^2_{IU}$ | IU Size | ELIR ESS (Info Unit) | ELIR ESS (Total Subjects) |
|---|---|---|---|---|
| 2:1 | 1.5 | 3 | 6 | 18 |
| 4:2 | 0.75 | 6 | 3 | 18 |
| 10:5 | 0.3 | 15 | 1.2 | 18 |

†: $a$ treatment subjects and $b$ control subjects

**Table 3**
*Prior ELIR ESS for the Risk Difference*

| $\theta_0$ | IU (Trt:Ctrl[†]) | IU Size | ELIR ESS (Info Unit) | ELIR ESS (Total Subjects) |
|---|---|---|---|---|
| 0.3 | 2:1 | 3 | 28.99 | 86.98 |
|  | 4:2 | 6 | 14.50 | 86.98 |
|  | 10:5 | 15 | 5.80 | 86.98 |
| 0.4 | 2:1 | 3 | 27.18 | 81.53 |
|  | 4:2 | 6 | 13.59 | 81.53 |
|  | 10:5 | 15 | 5.44 | 81.53 |

[†]: $a$ treatment subjects and $b$ control subjects

**Table 4**
*Prior ELIR ESS for the Log-Odds Ratio*

| $(\theta_0, s^2)$ | IU (Trt:Ctrl†) | IU Size | ELIR ESS (Info Unit) | ELIR ESS (Total Subjects) |
|---|---|---|---|---|
| $(0.4, 0.5^2)$ | 2:1 | 3 | 30.97 | 92.92 |
| | 4:2 | 6 | 15.49 | 92.92 |
| | 10:5 | 15 | 6.19 | 92.92 |
| $(0, 1^2)$ | 2:1 | 3 | 8.43 | 25.29 |
| | 4:2 | 6 | 4.21 | 25.29 |
| | 10:5 | 15 | 1.69 | 25.29 |

†: $a$ treatment subjects and $b$ control subjects

**Table 5**
*Predictive Consistency Check (Upper: Risk Difference; Lower: Odds Ratio)*

### Risk Difference

| Current trial SS (Trt:Ctrl†) | Prior ESS | Avg(Posterior ESS) | Avg(Posterior ESS) - Prior ESS - current SS |
|---|---|---|---|
| 100:50 | 81.5 | 309.7 | 78.2 |
| 200:100 | | 465.6 | 84.1 |
| 1000:500 | | 1668.4 | 86.9 |
| 10000:5000 | | 15169.7 | 88.1 |

### Odds Ratio

| Current trial SS (Trt:Ctrl†) | Prior ESS | Avg(Posterior ESS) | Avg(Posterior ESS) - Prior ESS - current SS |
|---|---|---|---|
| 100:50 | 92.9 | 227.8 | -15.1 |
| 200:100 | | 376.4 | -16.5 |
| 1000:500 | | 1575.4 | -17.5 |
| 10000:5000 | | 15075.2 | -17.7 |

†: $a$ treatment subjects and $b$ control subjects

**Table 6**
*Change-of-Variable Derivations for Binomial Endpoints*

| Effect Measure | $l_0 = h_1(p_0, p_1)$ | $\theta = h_2(p_0, p_1)$ | Jacobian $|J|$ |
|---|---|---|---|
| Risk difference | $\log\{\frac{p_0}{1-p_0}\}$ | $p_1 - p_0$ | $\frac{1}{p_0(1-p_0)}$ |
| Log-odds ratio | $\log\{\frac{p_0}{1-p_0}\}$ | $\log\{\frac{p_1}{1-p_1}\} - \log\{\frac{p_0}{1-p_0}\}$ | $\frac{1}{p_0(1-p_0)} \cdot \frac{1}{p_1(1-p_1)}$ |
| Log-rate ratio | $\log\{\frac{p_0}{1-p_0}\}$ | $\log\{p_1/p_0\}$ | $\frac{1}{p_1 p_0(1-p_0)}$ |